\documentclass[journal]{IEEEtran}

\ifCLASSINFOpdf
\else
   \usepackage[dvips]{graphicx}
\fi
\usepackage{url}

\usepackage{graphicx}
\usepackage{subfig}

\usepackage{amsmath,amsthm,amssymb}

\newcommand{\norm}[1]{\left\lVert#1\right\rVert}
\DeclareMathOperator*{\argmin}{arg\,min}
\DeclareMathOperator*{\argmax}{arg\,max}
\newtheorem{theorem}{Theorem}
\usepackage[ruled, noline, linesnumberedhidden, shortend]{algorithm2e}
\usepackage{mathrsfs}
\usepackage{cite}
\usepackage{bm}

\usepackage{framed}

\begin{document}

\title{A Fast and Scalable Polyatomic Frank-Wolfe Algorithm for the LASSO}

\author{Adrian Jarret, Julien Fageot, Matthieu Simeoni, \IEEEmembership{Member, IEEE}
\thanks{First submitted on December 4th, 2021, revised on January 31st, 2022.}
\thanks{This work was funded by the Swiss National Science Fundation (SNSF) under grants CRSII5\_193826  AstroSignals (A. Jarret and M. Simeoni), 200 021 181 978/1, “SESAM - Sensing and Sampling: Theory and Algorithms” (M. Simeoni) and P400P2\_194364 (J. Fageot).}
\thanks{A. Jarret and J. Fageot are from the School of Computer and Communication Sciences and M. Simeoni is from EPFL Center for Imaging, both at École Polytehnique Fédérale de Lausanne, Ecublens, Switzerland (email: name.surname@epfl.ch).}}

\maketitle

\begin{abstract}
We propose a fast and scalable \emph{polyatomic Frank-Wolfe (P-FW)}  algorithm for the resolution of high-dimensional LASSO regression problems. This algorithm improves upon traditional Frank-Wolfe methods by considering generalized greedy steps with \emph{polyatomic} (\textit{i.e.} linear combinations of multiple atoms) update directions, hence allowing for a more efficient exploration of the search space. To preserve sparsity of the intermediate iterates, we re-optimize the LASSO problem over the set of selected atoms at each iteration. For efficiency reasons, the accuracy of this re-optimization step is relatively low for early iterations and gradually increases with the iteration count. We provide convergence guarantees for our algorithm and validate it in simulated compressed sensing setups. Our experiments reveal that P-FW outperforms state-of-the-art methods in terms of runtime, both for FW methods and optimal first-order proximal gradient methods such as the \emph{Fast Iterative Soft-Thresholding Algorithm (FISTA)}.
\end{abstract}


\IEEEpeerreviewmaketitle

\section{Introduction}
\label{sec:intro}

\IEEEPARstart{L}{inear} regression is at the core of many inference tasks in signal processing, statistics or machine learning. With the advent of computing power and big data, modern regression problems tend to involve a large number of features, from a few thousands to hundreds of millions depending on the applications. Such large-scale regression problems are typically encountered in computational imaging setups, as for instance interferometric imaging in radio astronomy \cite{wiaux_compressed_2009, pan_leap_2017, simeoni2015towards, simeoni2021siml} and acoustics \cite{simeoni2019deepwave}, environmental monitoring~\cite{simeoni2021functional}, or medical imaging \cite{ravishankar_image_2020,8636261, roquette2018functional}. In such high dimensional settings, the LASSO regression problem \cite{tibshirani_regression_1996, tibshirani_lasso_2013} (also known as \emph{Basis Pursuit Denoising} in the signal processing community \cite{foucart2013invitation, chen_atomic_2001}) has received particular attention from the research community \cite{tibshirani2011regression, frandi_fast_2015}. Thanks to an $\ell_1$ regularization term, the LASSO promotes sparse estimates (\textit{i.e.} with relatively few active degrees-of-freedom) allowing for simpler and more interpretable models. The LASSO is most often used  in its penalized form:

\begin{equation}
    \label{eq:lasso}
    \argmin_{\mathbf{x} \in \mathbb{R}^N}{\norm{\mathbf{y} - \mathbf{Ax}}^2_2 + \lambda \norm{\mathbf{x}}_1},
\end{equation}
where $\mathbf{y}\in\mathbb{R}^L$ is a vector of observed values to be fitted, $\mathbf{A}\in\mathbb{R}^{L\times N}$ is the so-called \emph{design matrix}, $\mathbf{x}\in\mathbb{R}^N$ are the unknown regression coefficients and $\lambda>0$ is the penalty strength. In compressed sensing setups~\cite{foucart2013invitation}, the number of measurements $L$ is typically much smaller than the dimension $N$ of the problem.
The solution set to the LASSO problem \eqref{eq:lasso} can be shown to be non-empty, and the closed convex hull of (at most) $L$-sparse extreme points; see for example \cite[Theorem 6]{unser_representer_2016}, \cite[Theorem 6.8]{simeoni2020functional} and \cite{unser2017splines,gupta2018continuous,fageot2020tv,unser2021unifying, boyer2019representer,bredies2020sparsity, badoual2018periodic,debarre2020sparsest} for generalizations. Furthermore, if the design matrix coefficients are drawn according to a continuous probability distribution, the LASSO solution is unique with probability one, and guaranteed to be at most $L$-sparse~\cite{tibshirani_lasso_2013,foucart_mathematical_2013}.

Numerical solvers for addressing \eqref{eq:lasso} can be classified into three main categories: 
\begin{itemize}
    \item \emph{Greedy methods.} These methods identify at each iteration the variable \emph{most correlated} with the residuals and mark it as active. The vector of regression coefficients $\mathbf{x}$ is then updated in a direction computed from the set of variables activated so far \cite{frandi_fast_2015}. Examples include the \emph{Least-Angle Regression (LARS)} algorithm \cite{efron_least_2004, turlach2005algorithms} as well as specialised \emph{Frank-Wolfe (FW) methods}\footnote{Frank-Wolfe  methods are also referred to as  \emph{Conditional Gradient (CG) methods} in some communities.} \cite{harchaoui2012conditional, frandi_fast_2015,shalev2010trading, denoyelle_sliding_2020,bredies_inverse_2013, bredies_linear_2021,gong_mpgl_2017,boyd_alternating_2015,Braun_Pokutta_Tu_Wright_2019}.
    \item \emph{Coordinate Descent (CD) methods.} These methods  cyclically select \emph{one component} of $\mathbf{x}$ at a time and update it individually via a line search~\cite{li_coordinate_2009}. A very efficient CD method for the LASSO was proposed in \cite{friedman_pathwise_2007}.
    \item \emph{Proximal Gradient Descent (PGD) methods.} These methods update \emph{all components} of $\mathbf{x}$ at each iteration via successive gradient and proximal steps \cite{parikh2014proximal}. Examples include the well-known \emph{Fast Iterative Soft-Thresholding Algorithm (FISTA)} and its variants \cite{beck_fast_2009, scheinberg2014fast, liang2021improving}.
\end{itemize}
Thanks to their limited update scope, LARS, CD and FW scale relatively well with the number of features $N$. Indeed, the intermediate iterates they produce are often very sparse, hence alleviating the memory and computation bottlenecks of high dimensional regression. This is especially true for FW methods, which often yield significantly sparser solutions than CD or LARS \cite{frandi_fast_2015}. These algorithms are however relatively slow to converge with high accuracy, due to suboptimal convergence rates of $\mathcal{O}(1/k)$ \cite{jaggi_revisiting_2013, denoyelle_sliding_2020, zhao2020homotopic}. This is in contrast with PGD methods which, although less scalable due to their dense intermediate iterates, can achieve first-order optimal convergence rates of $\mathcal{O}(1/{k^2})$ when used in conjunction with  Nesterov's acceleration schemes \cite{beck_fast_2009, chambolle_convergence_2015, liang2021improving}.

In this work, we accelerate the FW method to make it competitive with optimal first-order proximal gradient descent methods such as FISTA. To this end, we propose generalized greedy steps, which we refer to as \emph{polyatomic} updates. Unlike standard greedy steps, the latter can activate multiple variables at the same time, allowing for a more efficient exploration of the search space. To ensure sparse intermediate iterates and further accelerate the algorithm, we also consider a partially-corrective step, which re-optimizes the LASSO problem over the set of previously activated atoms at each iteration. To minimize the computational cost of each iteration, this re-optimization step is performed with reduced, but gradually increasing, accuracy. The resulting algorithm is called \emph{Polyatomic Frank-Wolfe (P-FW)}. We provide convergence guarantees for the latter, and compare it to state-of-the-art FW methods and FISTA in simulated compressed sensing setups. In all investigated setups P-FW converges the fastest, sometimes by a very significant margin (\textit{e.g.} $\sim$4 times faster than FISTA, $\sim$20 times faster than FW methods).

The rest of this article is organized as follows. In Sections \ref{sec:frankwolfe} and \ref{sec:fw_for_lasso}, we  review the classical Frank-Wolfe algorithm and some of its variants, first in a generic setting and then more specifically for the LASSO problem. In Section \ref{sec:pfw}, we derive the  P-FW algorithm and establish  its convergence. In Section \ref{sec:experiments}, we test the performance of P-FW on simulated data.

\section{Frank-Wolfe and its Variants}
\label{sec:frankwolfe}
 
\subsection{Vanilla Frank-Wolfe}
\label{sec:vfw}

The Frank-Wolfe algorithm, as presented in 1956 \cite{frank_algorithm_1956}, aims at minimizing a general constrained convex optimization problem of the form 
\begin{equation}
    \label{eq:obj_fw}
    \argmin_{\mathbf{x} \in \mathcal{D}}f(\mathbf{x})
\end{equation}
where $f$ is a convex and continuously differentiable cost function and the domain $\mathcal{D}$ is a compact convex subset of a some Banach space. The vanilla version of the procedure is provided in Algorithm \ref{alg:vfw} below.

\begin{algorithm}[h!]
\DontPrintSemicolon
\caption{Vanilla Frank-Wolfe Algorithm (V-FW)}
\label{alg:vfw}
Initialize $\mathbf{x}_0 \in \mathcal{D}$ \;
\For{$k=1, 2 \cdots$}{
    \nlset{1)\hspace{-3pt}} \label{algstep:gradient}Find an update direction:\\
    \ \ $\mathbf{s}_k \in \argmin_{\mathbf{s}\in\mathcal{D}}{ \langle \nabla f(\mathbf{x}_k), \mathbf{s} \rangle }$ \;

    \nlset{2.a)\hspace{-3pt}} \label{algstep:step_size}Step size: $\gamma_k \gets \frac{2}{k+2}$\;
    \nlset{2.b)\hspace{-3pt}} \label{algstep:reweighting}Reweight: $\mathbf{x}_{k+1} \gets (1-\gamma_k) \mathbf{x}_k + \gamma_k \mathbf{s}_k $\;
}
\end{algorithm}

The update direction search in step \ref{algstep:gradient}\hspace*{3pt} consists in minimizing a linear function over a convex set, which leads to solutions lying on the set's boundary. When the domain $\mathcal{D}$ is a polytope, the minimum is necessarily reached for one of its extreme points or \emph{atoms}, hence allowing to restrict step~\ref{algstep:gradient}\hspace*{3pt} to the generating atoms of $\mathcal{D}$ only. The computation of the update direction is very cheap since it is projection-free. This is in contrast with \emph{Projected Gradient Descent} methods for \eqref{eq:obj_fw}, which could involve very expensive projections onto the set $\mathcal{D}$ at each iteration.

\subsection{Known Frank-Wolfe variants}
\label{sec:fwvariants}

Several variants of the FW algorithm have been proposed in the literature. Adopting the same formalism as in \cite{jaggi_revisiting_2013}, we review some of these variants that we will later rely on.
\subsubsection{Exact Line Search}
It is possible to replace step \ref{algstep:step_size}\hspace*{3pt} by an exact line search, leading to Algorithm \ref{alg:els_fw}.
\begin{algorithm}[h!]
\DontPrintSemicolon
\caption{Exact Line Search}
\label{alg:els_fw}
\emph{... Same as Algorithm \ref{alg:vfw}, replacing steps \ref{algstep:step_size}\hspace*{3pt} by:} \;
    \nlset{\ \ 2.a')\hspace{-3pt}} \label{algstep:els} Solve: $ \gamma_k \gets \argmin_{\gamma \in [0, 1]} f((1-\gamma)\mathbf{x}_k + \gamma \mathbf{s}_k)$
\end{algorithm}

This step is typically performed when a closed-form solution for step \ref{algstep:els}\hspace*{3pt} exists, \textit{e.g.} with quadratic objective functionals.

\subsubsection{Approximate Linear Subproblem}

In cases where the gradient minimization step \ref{algstep:gradient}\hspace*{3pt} is expensive (\textit{e.g.} complex domain shape), the latter can be performed approximately. This yields Algorithm \ref{alg:als_fw},  where $\delta > 0$ controls the approximation quality: 

\begin{algorithm}[h!]
\DontPrintSemicolon
\caption{Approximate Linear Subproblem of quality $\delta > 0$}
\label{alg:als_fw}
\emph{... Same as Algorithm \ref{alg:vfw} or \ref{alg:els_fw}, replacing step \ref{algstep:gradient}\hspace*{3pt} by:} \;
    \nlset{1')} \label{algstep:approx_linear} Find: $\mathbf{s}_k \in \mathcal{D}$ such that $\langle \nabla f(\mathbf{x}_k), \mathbf{s}_k \rangle \geq \min_{\mathbf{s} \in \mathcal{D}} \langle \nabla f(\mathbf{x}_k), \mathbf{s} \rangle + \gamma_k \delta$ \\ with $\gamma_k = {2}/{(k+2)}$ \;
\end{algorithm}

\subsubsection{Fully-Corrective FW}

Another commonly used variant is known as \emph{Fully-Corrective FW} (FC-FW), which re-optimizes $f$ over the convex-hull of all previously selected atoms at each iteration. This results in Algorithm \ref{alg:fc_fw} below, where $\mathrm{Conv}$ denotes the convex hull of a given set.
\begin{algorithm}[h!]
\DontPrintSemicolon
\caption{Fully-Corrective Variant (FC-FW)}
\label{alg:fc_fw}
\emph{... Same as Algorithm \ref{alg:vfw} or \ref{alg:als_fw}, replacing steps \ref{algstep:step_size}\hspace*{3pt}/\ref{algstep:reweighting}\hspace*{3pt} by:} \;
    \nlset{2')} \label{algstep:fully_corrective} Solve: $ \mathbf{x}_{k+1} \gets \argmin_{\mathbf{x} \in \mathrm{Conv}(\mathbf{s}_1, \dots, \mathbf{s}_k)} f(\mathbf{x})$ \;
\end{algorithm}

The re-optimization step \ref{algstep:fully_corrective} allows for more progress to be made at each iteration and can potentially discard wrongly selected atoms. This procedure however requires solving at each iteration an optimization problem that may turn out to be as difficult as the original one, and thus computationally expensive. 

\subsection{Convergence guarantees}
Theorem 1 of \cite{jaggi_revisiting_2013} shows that the sequence of iterates $\mathbf{x}_k$ produced by FW or its variants achieves a convergence rate of $\mathcal{O}(1/k)$ in terms of the {objective functional}. More specifically, let $C_f>0$ denote the \emph{curvature constant}\footnote{See \cite{jaggi_revisiting_2013} for a definition.} of $f$, and $f^*$ its optimal value, the iterates of Algorithms \ref{alg:vfw} to \ref{alg:fc_fw} satisfy
\begin{equation}
    f(\mathbf{x}_k) - f^* \leq \frac{2}{k+2}(C_f + 2\delta), \qquad \forall k \geq 1,
\end{equation}
with $\delta = 0$ for Algorithms \ref{alg:vfw}, \ref{alg:els_fw}, and potentially \ref{alg:fc_fw} if the latter does not perform the optional approximate gradient minimization step.

\section{Frank-Wolfe for the LASSO}
\label{sec:fw_for_lasso}
While not of the form \eqref{eq:obj_fw} due to its non-differentiable cost functional,  the penalized LASSO regression problem \eqref{eq:lasso} can still be brought in the scope of the FW algorithm by means of an \emph{epigraphical lift}, as proposed in~\cite{harchaoui2012conditional,denoyelle_sliding_2020}. To this end, we introduce the bounded norm cone $C = \left\{(t, \mathbf{x}) \in \mathbb{R}_+ \times \mathbb{R}^N : \norm{\mathbf{x}}_1 \leq t \leq M \right\}$ with $M = \norm{\mathbf{y}}^2_2 / 2\lambda$. With similar arguments as in  \cite[Lemma 4]{denoyelle_sliding_2020}, it is then possible to show that  \eqref{eq:lasso} is equivalent to the constrained optimization problem:
\begin{equation}
    \label{eq:diff_lasso}
    \argmin_{(t, \mathbf{x}) \in C}{\norm{\mathbf{y} - \mathbf{Ax}}^2_2 + \lambda t}.
\end{equation}
Observe that \eqref{eq:diff_lasso} is indeed of the form \eqref{eq:obj_fw}, since the objective function $f(t, \mathbf{x})=\norm{\mathbf{y} - \mathbf{Ax}}^2_2 + \lambda t$ is continuously differentiable and $C \subset \mathbb{R}_+ \times \mathbb{R}^N$ is compact and convex. Applied to \eqref{eq:diff_lasso}, the update direction step \ref{algstep:gradient}\hspace*{3pt} in Algorithm~\ref{alg:vfw}  becomes 
\begin{equation}
    \label{eq:direction_search}
    (t_k, \mathbf{s}_k) \in \argmin_{(t,\mathbf{x})\in C}{ \langle \bm{\eta}_k, \mathbf{x} \rangle + \lambda t} 
\end{equation}
where $\bm{\eta}_k = \frac{1}{\lambda}\mathbf{A}^T(\mathbf{y} - \mathbf{Ax}_k)$ is the so-called \emph{empirical dual certificate} at iteration $k$  \cite{denoyelle_sliding_2020}. Since the extreme points of $C$ are easily shown to be of the form $(M, \pm M \mathbf{e}_{i_k})$ (with $\mathbf{e}_{j}$ the $j$-th canonical basis vector), solving \eqref{eq:direction_search} is equivalent to finding the index $i_k$ such that
\begin{equation}
    \label{eq:update_index}
    i_k = \argmax_{i \in \{1, \dots, N\}} |(\bm{\eta}_k)_i|.
\end{equation}
Note that the lift variable $t$ is constant across iterations and is thus omitted in what follows. To sum up, the update direction step \ref{algstep:gradient}\hspace*{3pt} amounts to the creation of a new atom $\pm M \mathbf{e}_{i_k}$ that will be reweighted and added to the current iterate $\mathbf{x}_k$. The selected index $i_k$ is associated to the column of $\mathbf{A}$ which is most correlated with the residuals $\mathbf{y}-\mathbf{A}\mathbf{x}$. Note that the exact same update step appears in the greedy \emph{Matching Pursuit (MP)} \cite{mallat_matching_1993} and \emph{Orthogonal Matching Pursuit (OMP)} \cite{pati_orthogonal_1993,mallat_adaptive_1994} algorithms, commonly used in signal processing for sparse recovery.

When considering the approximated subproblem variant, step \ref{algstep:approx_linear} of Algorithm \ref{alg:als_fw} becomes
\begin{align}
    \label{eq:threshold_certificate}
    &\text{Find} \; i_k \in \{1, \dots, N\}, \; \text{such that}\nonumber\\
    &|(\bm{\eta}_k)_{i_k}| \geq \max_{i \in \{1, \dots, N\}} |(\bm{\eta}_k)_i| - \gamma_k \delta = \norm{\bm{\eta}_k}_\infty - \gamma_k \delta.
\end{align}
Moreover, the fully-corrective step \ref{algstep:fully_corrective} of Algorithm \ref{alg:fc_fw} becomes: 
\begin{align}
    &\tilde{\mathbf{x}}_{k+1} \gets \argmin_{\tilde{\mathbf{x}} \in \mathbb{R}^{\mathrm{Card}(\mathcal{S}_{k})} }{ \norm{\mathbf{y} - \mathbf{A}_{\mathcal{S}_{k}}\tilde{\mathbf{x}}}^2_2 + \lambda \norm{\tilde{\mathbf{x}}}_1}\label{eq:fc_for_fw}\\
    & \mathbf{x}_{k+1}[\mathcal{S}_{k}] \gets \tilde{\mathbf{x}}_{k+1}, \nonumber \\
    & \mathbf{x}_{k+1}[\overline{\mathcal{S}_{k}}] \gets \bm{0}, \nonumber
\end{align}
where $\mathcal{S}_k := \{i_1, \dots, i_k\} \subset \{1, \dots, N\}$  denotes the set of the active indices up to iteration $k$ and $\mathbf{A}_{\mathcal{S}_k} \in \mathbb{R}^{L \times \mathrm{Card}(\mathcal{S}_k)}$ denotes the restriction of $\mathbf{A}$ to its columns with indices in $\mathcal{S}_k$.
Note that in comparison with the initial problem \eqref{eq:lasso}, the dimension of \eqref{eq:fc_for_fw} is dramatically reduced, from $N$ to $\mathrm{Card}(\mathcal{S}_k)$ at iteration $k$.

\section{Polyatomic Frank-Wolfe}
\label{sec:pfw}
Observe that there might exist multiple indices that simultaneously satisfy the approximate gradient minimization step~\eqref{eq:threshold_certificate}, each yielding different but equally valid atomic update directions. Unlike state-of-the-art approaches which limit themselves to selecting one of these directions arbitrarily, we propose taking \emph{polyatomic} update directions (\textit{i.e.} expanding the set of active indices with multiple indices at once). This is indeed possible since any convex combination of the potential atomic update directions is also a solution to  \eqref{eq:threshold_certificate} --although in general not on the boundary of the constraint convex set $C$ anymore. The idea behind such polyatomic update directions is to accelerate the search space exploration, and consequently the convergence of the algorithm. Polyatomic update steps have already been used in other contexts (see for example \cite{donoho_sparse_2012} for OMP), but to the best of our knowledge, its combination with a FW-type algorithm has not yet been explored.

The main difficulty in working with polyatomic update directions consists in correctly estimating the weights associated to each atom in the convex combination. When a fully-corrective step is additionally performed, choosing the convex optimization weights optimally is no longer a necessity since \eqref{eq:fc_for_fw} will anyway re-optimize these weights. We therefore propose to use suboptimal polyatomic update steps with arbitrary weights (\textit{e.g.} the mean of all atoms) followed by fully-corrective steps. In practice, to further reduce the computational cost of each iteration, we also propose to prematurely stop the fully-corrective steps, resulting in \emph{partially-corrective} steps. Thanks to an adaptive stopping criterion, these steps are designed to be more and more accurate as the number of iterations increases. Indeed, early partially-corrective steps need not be very accurate since the iterates evolve a lot from one iteration to the other. 

Putting all these ingredients together results in Algorithm~\ref{alg:pfw}, which we call the \emph{Polyatomic Frank-Wolfe (P-FW)} algorithm. The partial-correction procedure is described in Algorithm \ref{alg:ista}. It consists in a warm-started iterative soft-thresholding algorithm with a stopping criterion based on the relative improvement of the iterates.
\begin{algorithm}[b!]
\DontPrintSemicolon
\caption{Polyatomic FW (P-FW) of quality $\delta > 0$}
\label{alg:pfw}
Initialize: $\mathbf{x}_0 \gets 0, \mathcal{S}_0 \gets \emptyset$ \;
\For{$k=1, 2 \cdots$}{
    $\gamma_k \gets 2/(k+2)$\;
    \nlset{1".a)\hspace{-5pt}}\label{algstep:multi}Polyatomic exploration:\;
    \qquad $\mathcal{I}_k = \left\{ 1 \leq j \leq N : |\bm{\eta}_k|_j \geq \norm{\bm{\eta}_k}_\infty - \delta\gamma_k\right\}$\;
    \qquad $\mathbf{s}_k \gets \left(\sum_{i \in \mathcal{I}_k} \mathbf{e}_i\right)/\ \text{Card}(\mathcal{I}_k)$\;
    \nlset{1".b)\hspace{-5pt}} Update active indices: $\mathcal{S}_{k} \gets \mathcal{S}_{k-1} \cup \mathcal{I}_k$\;
    \nlset{2".a)\hspace{-5pt}} Set accuracy threshold: $\varepsilon_{k} = \varepsilon_0\gamma_k$\;
    \nlset{2".b)\hspace{-5pt}}\label{algstep:ista}Update active weights:\;
    \quad $\mathbf{x}_{k+1/2}\gets (1-\gamma_k)\mathbf{x}_k + \gamma_k \mathbf{s}_k$\;
    \quad $\mathbf{x}_{k+1} \gets \mathtt{partial\_correction}(\mathbf{x}_{k+1/2}, \mathcal{S}_{k}, \varepsilon_{k})$ \;
    $k \gets k+1$\;
}
\end{algorithm}

\begin{algorithm}[t!]
\DontPrintSemicolon
\caption{\texttt{partial\_correction}$(\mathbf{x}, \mathcal{S}, \varepsilon)$}
\label{alg:ista}
Initialize: $\tilde{\mathbf{u}}_0 \gets \mathbf{x}[\mathcal{S}], k\gets 1, \tau \gets 1/ \norm{\mathbf{A}_\mathcal{S}}_2$\;
\While{$\norm{\tilde{\mathbf{u}}_k - \tilde{\mathbf{u}}_{k-1}}_2 > \varepsilon \norm{\tilde{\mathbf{u}}_{k-1}}_2$}{
    $\tilde{\mathbf{v}}_{k+1} = {\tilde{\mathbf{u}}}_k - \tau \mathbf{A}_\mathcal{S}^T(\mathbf{A}_\mathcal{S} \tilde{\mathbf{u}}_k - \mathbf{y})$\;
    $\tilde{\mathbf{u}}_{k+1} = \mathrm{sgn}(\tilde{\mathbf{v}}_{k+1}) \mathrm{max}(\boldsymbol{0}, \lvert\tilde{\mathbf{v}}_{k+1}\rvert - \tau\lambda)$\;
    $k \gets k+1$\;
}
Return: $\mathbf{u} \in \mathbb{R}^N$ with $\mathbf{u}[\mathcal{S}] = \tilde{\mathbf{u}}_k$, $\mathbf{u}[\overline{\mathcal{S}}] = \mathbf{0}$\;
\end{algorithm}


Theorem \ref{th:cvg} proves the convergence of P-FW, relying on \cite[Theorem 1]{jaggi_revisiting_2013}.
\begin{theorem}[Convergence of Polyatomic FW]
    \label{th:cvg}
    Consider Algorithm \ref{alg:pfw} with parameter $\delta > 0$. Denote  by $\mathscr{L}(\mathbf{x}) = \norm{\mathbf{y} - \mathbf{Ax}}^2_2 + \lambda \norm{\mathbf{x}}_1$ the LASSO cost function and $\mathscr{L}^*$ its minimal value. For each $k \geq 1$, the iterates $\mathbf{x}_k$ satisfy
    \begin{equation*}
        \mathscr{L}(\mathbf{x}_k) - \mathscr{L}^* \leq \frac{2}{k+2}(C_f + 2\delta)
    \end{equation*}
    where $C_f$ is the {curvature constant} of \eqref{eq:diff_lasso}.
\end{theorem}
\begin{proof}
    First, we remark that Lemma 5 of \cite{jaggi_revisiting_2013} holds for $\mathbf{s}_k$ being any convex combination of solutions of \eqref{eq:threshold_certificate}. Then we only need $\mathscr{L}(\mathbf{x}_{k+1}) \leq \mathscr{L}(\mathbf{x}_{k+1/2})$ to apply the proof of \cite{jaggi_revisiting_2013}. This is necessarily the case since  the partial correction solver ISTA is monotonic and initialized with $\mathbf{x}_{k+1/2}$.
\end{proof}

Note that Theorem \ref{th:cvg} only provides an upper bound on the  convergence rate of Algorithm \ref{alg:pfw}. In practice, the algorithm tends to converge much faster, as illustrated in the next section.

\section{Numerical Simulations}
\label{sec:experiments}
\begin{figure}[t!]
\begin{center}
\vspace*{-.5cm}
\subfloat[$K=32, \alpha = 16$]{
  \includegraphics[trim={0 15 0 0},width=.48\linewidth, clip]{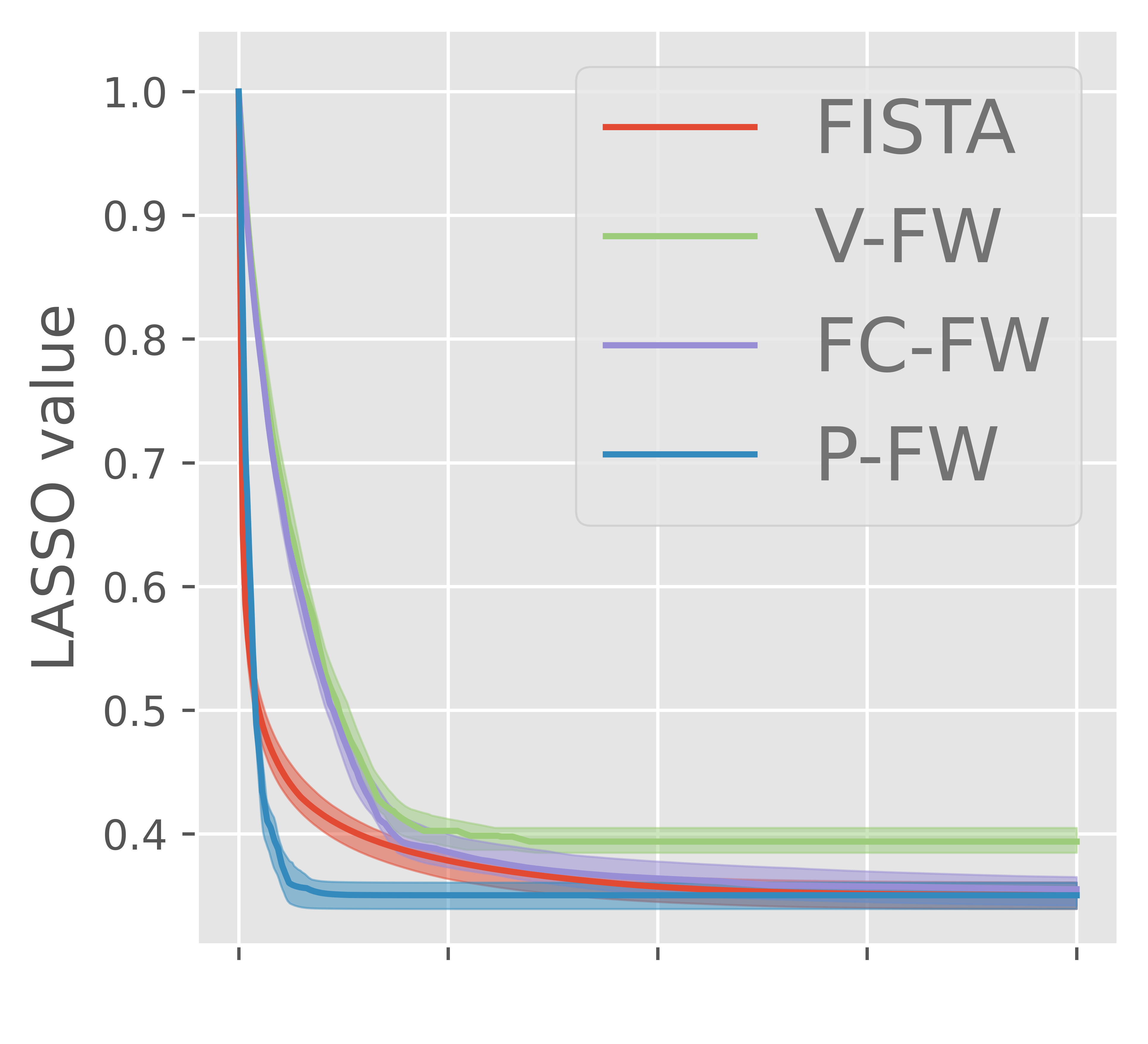}
  }\hfill
\subfloat[$K=32, \alpha = 64$]{
  \includegraphics[trim={0 15 0 0},width=.45\linewidth, clip]{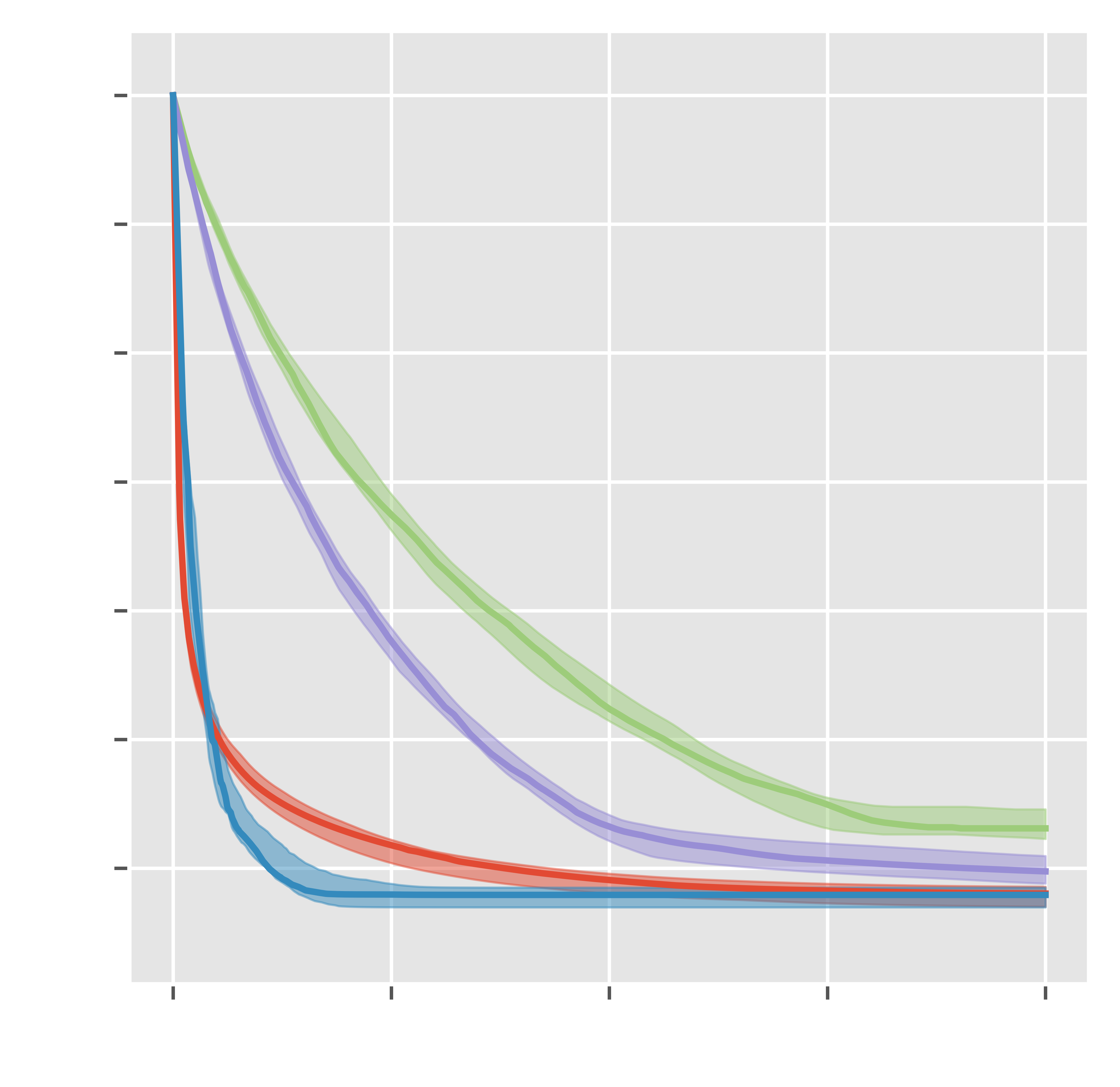}
  }
  
\subfloat[$K=64, \alpha = 16$]{
  \includegraphics[trim={0 15 0 0},width=.48\linewidth, clip]{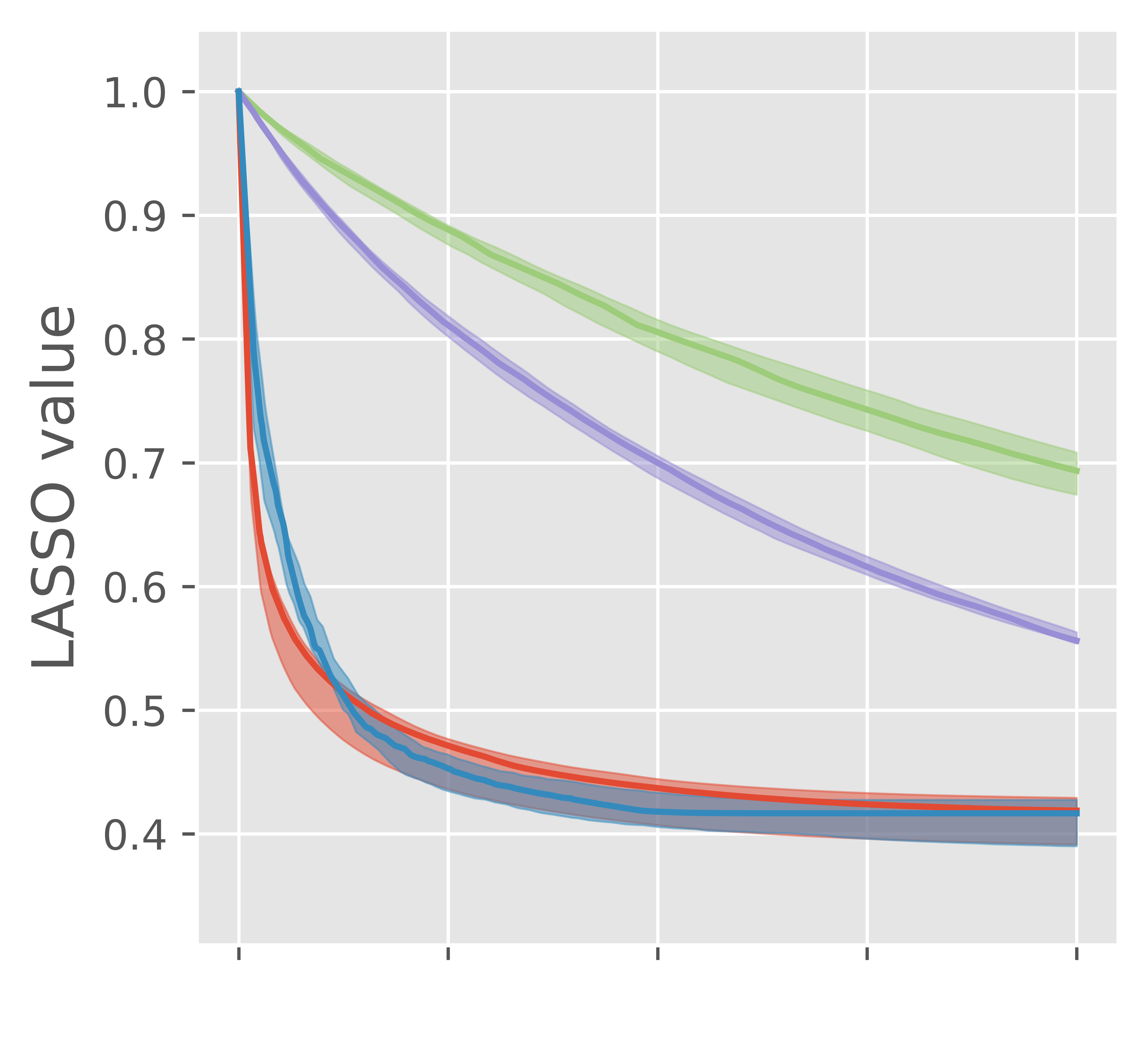}
  }\hfill
\subfloat[$K=64, \alpha = 64$]{
  \includegraphics[trim={0 15 0 0},width=.45\linewidth, clip]{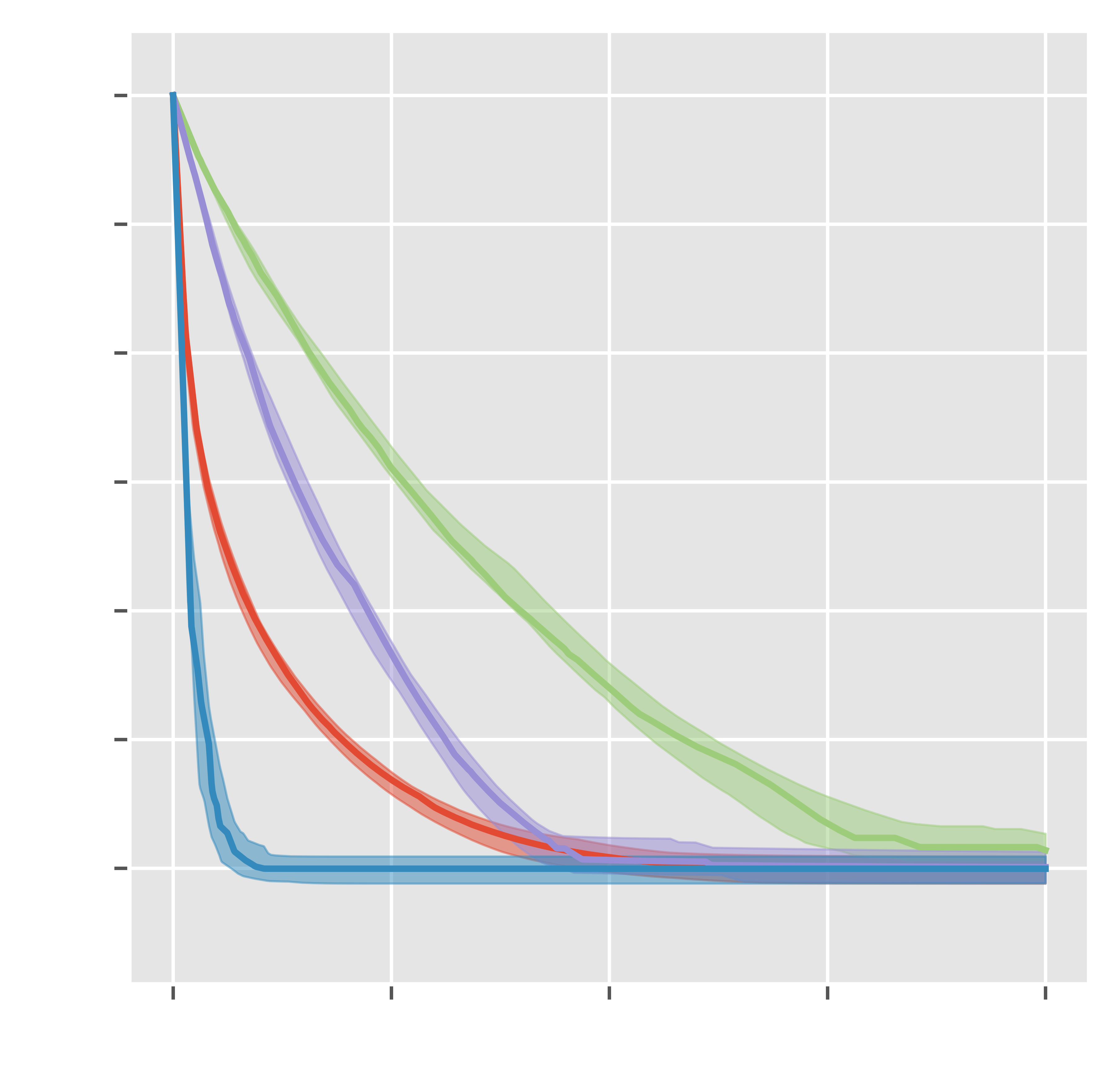}
  }
  
\subfloat[$K=128, \alpha = 16$ ]{
  \includegraphics[width=.48\linewidth]{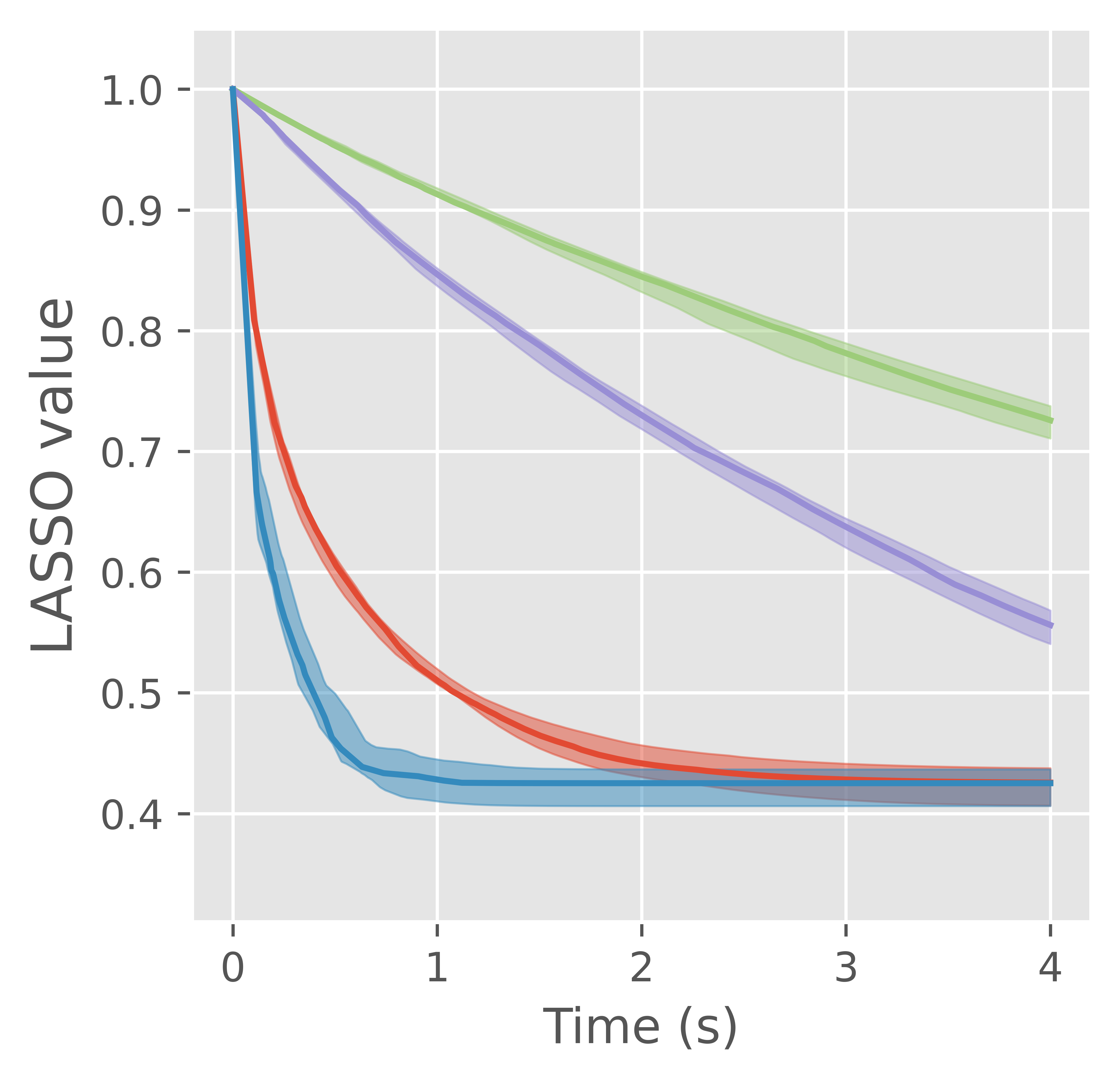}
  }\hfill
\subfloat[$K=128, \alpha = 64$]{
  \includegraphics[width=.45\linewidth]{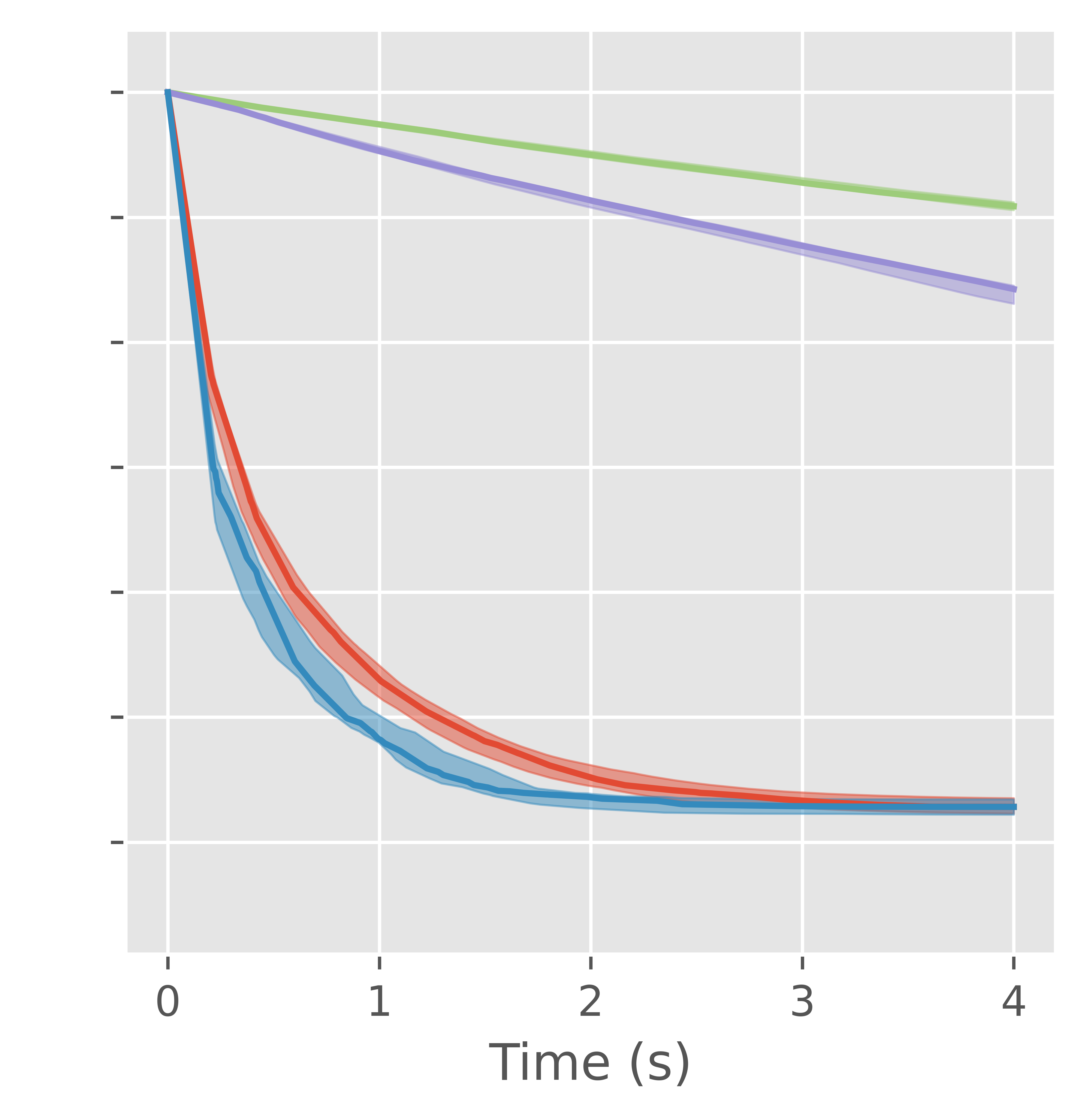}
  }
\caption{Value of the LASSO objective functional across reconstruction time for the four algorithms benchmarked (FISTA, V-FW, FC-FW, P-FW) and various values of $K$ and $\alpha$. The timings are for a ThinkPad T14, Intel Core i7 (4C/8T) @ 1.8GHz with 32GB RAM. These results can be reproduced using the Python scripts provided in our GitHub repository~\cite{adriaj}.}
\label{fig:plots}
\end{center}
\end{figure}

For our simulations, we consider the following data model:
\begin{equation}
    \label{eq:model}
    \mathbf{y} = \mathbf{Ax}_0 + \mathbf{w},
\end{equation}
where $\mathbf{x}_0 \in \mathbb{R}^N$ is a $k$-sparse vector of dimension $N = 16384$, $\mathbf{A} \in \mathbb{R}^{L\times N}$ is a Gaussian random matrix and $\mathbf{w}$ an additive Gaussian white noise with a PSNR of $20$ dB. We set the number of measurements to $L = \alpha K \ll N$ for some oversampling factor $\alpha >1$. We then solve a LASSO problem \eqref{eq:lasso} with input data $\mathbf{y}$ with four algorithms\footnote{Our implementation  is based on the \texttt{Pycsou} optimization package \cite{SIMEONI_Pla_2021}.}: \emph{FISTA},  \emph{Vanilla FW} (V-FW) (with optimal line search, Algorithm \ref{alg:els_fw}), \emph{Fully-Corrective FW} (FC-FW) Algorithm \ref{alg:fc_fw}, and our proposed \emph{Polyatomic FW} (P-FW). 
We let each algorithm run for a maximum of $4$ seconds and monitor the evolution of the LASSO objective functional value across reconstruction time. We perform this experiment for various values of $K$ and $\alpha$, each time re-running the algorithms for  $15$ independent noise realizations. 
Fig.~\ref{fig:plots} shows the median LASSO objective functional value vs time, as well as the interquartile distance (shaded).

Although FC-FW is slightly faster than V-FW, these two algorithms are significantly slower to converge than P-FW or FISTA. Remarkably, P-FW is able to  outperform the  optimal first-order method FISTA in most setups. The most striking example comes from configuration (d) of Fig.~\ref{fig:plots}, in which FISTA takes on average more than $2.0$s to reach an objective functional value reached by P-FW in less than $0.5$s ($\sim$4$\times$ speedup). Furthermore, we remark that the performances of V-FW and FC-FW degrade with larger values of $K$ (ground truth $\mathbf{x}_0$ less sparse), whereas this is not the case for P-FW. It is able to remain competitive with FISTA, even in the less sparse setups ($K=128$) of plots (e) and (f). 

\section{Conclusion}
\label{sec:print}

In this paper, we introduce a novel polyatomic FW algorithm, tailored to large-scale LASSO problems. Our procedure leverages polyatomic update directions and partially-corrective re-optimization steps for fast convergence. By preserving sparse intermediate iterates, the algorithm is well-suited for high dimensional regression problems. As revealed by our numerical simulations, the algorithm converges faster than standard state-of-the-art LASSO solvers (from 1.5 to 20 times faster depending on the setup and algorithm), notably including the optimal first-order method FISTA.
Future work will include the application of this promising algorithm to the challenging problems of sparse radio-interferometric imaging~\cite{wiaux_compressed_2009}. \vspace{-1.5em}
{\let\thefootnote\relax\footnote{{\textbf{Acknowledgement.} The authors are grateful to Martin Vetterli for his expert advice as well as Quentin Denoyelle and Eric Bezzam for their useful feedback and/or proofreading on earlier versions of this work.}}}


\pagebreak
\bibliographystyle{ieeetr}
\bibliography{refs}

\end{document}